# Reactivity and Survivability of Glycolaldehyde in Simulated Meteorite Impact Experiments

V.P. McCaffrey[1*], N.E.B. Zellner[2], C.M. Waun[1], E.R. Bennett[1], and E.K. Earl[1]

[1]Department of Chemistry, Albion College, Albion, MI 49224, USA

[2]Department of Physics, Albion College, Albion, MI 49224, USA

*vmccaffrey@albion.edu, 517-629-0622 (Phone), 517-629-0264 (Fax)

**Abstract**: Sugars of extraterrestrial origin have been observed in the interstellar medium (ISM), in at least one comet spectrum, and in several carbonaceous chondritic meteorites that have been recovered from the surface of the Earth. The origins of these sugars within the meteorites have been debated. To explore the possibility that sugars could be generated during shock events, this paper reports on the results of the first laboratory impact experiments wherein glycolaldehyde, found in the ISM, as well as glycolaldehyde mixed with montmorillonite clay, have been subjected to reverberated shocks from ~5 GPa to >25 GPa. New biologically relevant molecules, including threose, erythrose and ethylene glycol, were identified in the resulting samples. These results show that sugar molecules can not only survive but also become more complex during impact delivery to planetary bodies.

**Introduction**

The discovery of biologically relevant molecules in extraterrestrial samples has spurred interest in understanding their formation mechanisms and their survivability upon delivery to a planet. Meteorites and comets are known to contain a rich array of prebiotic compounds, and studies have focused mainly on the identification of the many classes of compounds that have been found in them. For example, many sugar alcohols and at least one sugar (Cooper et al. 2001) and ~70 natural and unnatural amino acids (Glavin et al. 2012; Burton et al. 2011; Botta and Bada 2002; Pizzarello et al. 1991) have been identified in carbonaceous chondrite meteorites. Another amino acid, glycine, has been identified in cometary dust returned by the *Stardust* mission (Elsila et al. 2009). The potential importance of meteorites and comets for delivering organic compounds to the early Earth that could have contributed to the origin of life has been one of the principal justifications for ESA's Rosetta (*in situ* analysis of comet 67P/Churyumov-Gerasimenko) and NASA's OSIRIS-REx (sample return from carbonaceous asteroid 101955 Bennu) missions. Current estimates of the total flux of extraterrestrial material range from 30 x $10^6$ kg/year (Love and Brownlee 1993) and even higher on Mars (Flynn and McKay 1990) today to 1 - $10^5$x the current flux on the early Earth (Owen 1998; Zahnle and Sleep 2006; Marty and Yokochi 2006; Pasek and Lauretta 2008). Thus, the effects of pressure and temperature on the chemistry of amino acids during and after shock, such as via impact delivery, are important to understand and have been investigated (Bertrand et al. 2009; Blank et al. 2001; Peterson et al. 1997). Impact studies have also looked at the chemistry of polycyclic aromatic hydrocarbons (Mimura and Toyama 2005) and the formation of amino acids from volatiles and water ice in the hypervelocity impact shock of typical comet ice mixtures (Martins et al. 2013).

Biologically interesting molecules have also been detected in the interstellar medium (ISM). Glycolaldehyde ($C_2(H_2O)_2$, GLA) has been observed in a variety of locations, including the molecular cloud Sagittarius B2N (e.g., Halfen et al. 2006; Hollis et al. 2004; Hollis et al. 2001; Hollis et al. 2000), the hot molecular core G31.41+0.31 (Beltrán et al. 2009), and the proto-star IRAS 16293-2422 (Jørgensen et al. 2012). With its with two carbon atoms, GLA is of considerable interest to astrobiology because it has been postulated to be a precursor in the synthesis of ribose in interstellar space (Jalbout et al. 2007), and it is an accepted precursor for studying the formose reaction under a variety of conditions (Kim et al. 2011; Lambert et al. 2010). A variety of formation mechanisms have been proposed for GLA, including its formation inside icy grain mantles that are photoprocessed by ultraviolet starlight (Jalbout et al. 2007; Bennet and Kaiser 2007; Sorrell 2001) or on grain surfaces (Woods et al. 2012).

Extraterrestrial organic materials could be significant sources of starting materials for the formation of life (Chyba et al. 1990; Anders 1989), and the sugar derivatives (Cooper et al. 2001) may have provided precursor materials for the formation of the backbones of DNA and RNA. The timing of the appearance of ribose, in particular, on the prebiotic Earth is a key element to understanding how life may have started. An alternate theory suggests that threose (a C4 sugar) could have been an early progenitor of RNA (e.g., Yu et al. 2012). These models underscore the importance of understanding the many formation pathways of simple sugars and their derivatives. For example, it has been proposed that the formose process might be responsible for the formation of sugars (Cooper et al. 2001; Jalbout et al. 2007; Kim et al. 2011). In some models, the sugars could have been synthesized *in situ* on an early Earth and in others, delivered on meteorites and micrometeorites (Cooper et al. 2001; Harman et al. 2013).

Montmorillonite clay, usually formed by the weathering of volcanic ash (Papke 1969; Joshi et al. 2009; Delano et al. 2010), has been shown to catalyze the formation of many different types of complex prebiotic molecules and has been the focus of many studies involving the formation of RNA oligomers (Joshi et al. 2009; Ferris 2005; Ferris 1998; Ferris and Ertem 1993; Ferris and Hagan 1986). The layered silicate structure of the clay provides both acidic and basic sites which give rise to the large number of reactions that can be catalyzed by these materials. The presence of clays in general is ubiquitous in the solar system. In fact, they have been observed on a wide variety of extraterrestrial objects, including meteorites (MacKinnon and Kaser 1988; Zolensky and Keller 1991; Keller and Zolensky 1991), interplanetary dust particles (Zolensky and Keller 1991; Keller and Zolensky 1991; Rietmeijer and MacKinnon 1985), comets (Wozniakiewics et al. 2010; Lisse et al. 2006), Europa (Shirley et al. 2013), and Mars (e.g., Sun and Milliken 2014;  Farley et al. 2014; Stephenson et al. 2013; Bishop et al. 2008; Wray et al. 2008).

In this work, we investigate the survivability and reactivity of glycolaldehyde (GLA) under conditions that mimic those of impacting objects. This paper reports the results of the first experiments subjecting glycolaldehyde (GLA) mixed with a montmorillonite-rich bentonite (hereafter GLA/clay) to impacts conducted with the flat-place accelerator at the NASA Johnson Space Center's Experimental Impact Laboratory. The simple sugar precursor, glycolaldehyde, was mixed with a mineral matrix and subjected to a series of impact events with pressures ranging from ~5 to >25 GPa, pressures that would be experienced by incoming extraterrestrial objects.

**Materials and Methods**

*Impact Experiments: GLA and Clay*

The materials were subjected to controlled shock stresses in the Experimental Impact Laboratory at the Johnson Space Center using well-established techniques (Hörz 1970; Gibbons and Ahrens 1971; Stöffler 1972).  Because the shock-reverberation method has been described in detail elsewhere (Gibbons 1974), an overview has been provided in the Supplemental Materials.

Experiments involved a 20:1 mass ratio of a montmorillonite (bentonite) clay to GLA. The clay was collected from Belle Fourche, SD and is known to be catalytic (Joshi et al. 2009). Before use, the clay was analyzed to determine organic content. A sample of the received clay was washed separately with pyridine and tetrahydrofuran (THF) and the resulting washes were treated with BSTFA as described below and analyzed by GC/MS. No residual organic material was found. As a result, the clay was used in the shock experiments with no processing. In all three experiments, dry GLA and clay were blended together, and ~100 mg (see Table S1 for complete details of the impact experiments) of the mixture were packed into each stainless steel sample well as tightly as possible to minimize porosity that could affect localized reverberations and associated stress concentrations (Peterson et al. 1997; Kieffer 1971). The chamber was evacuated to below 200 mTorr and the target was then impacted by either a Lexan projectile with no flyer plate, an aluminum (Al 2024) flyer plate, or a stainless steel (SS 304) flyer plate, depending on the desired shock stress. Given the composition of the projectiles and the flyer plates, with velocities averaging ~1.1 km/s, the samples experienced shock pressures of 4.65 GPa, 12 GPa and 25 GPa.

*Shocked Materials: GLA and Clay*
*Physical Characteristics*

Samples of shocked GLA and clay proved robust under the range of pressures and at least 94% of the sample was recovered in all experiments. Before impact, the samples were a uniform grey powder; when the target assemblies were opened after the impact event, the samples were recovered as compacted disks. At the lowest pressure (4.65 GPa, shot 3598), there was a slight darkening of the sample surface, but when the surface was scratched, the interior of the disk was a lighter uniform grey color. At 12 GPa (shot 3604), a viscous liquid was seen on the surface of the recovered sample disk. The liquid dried quickly and the residue was included in the sample analysis. The sample of the highest pressure shot with GLA/clay (25.1 GPa, shot 3622) was recovered as a dark grey disk.

*Chemical Analysis*

In a typical analysis, ~25 mg of shocked sample (corresponding to 1.2 mg of GLA) were placed in a small scintillation vial. The organic materials were extracted from the clay by addition of 1 mL of THF. The mixture was sonicated to ensure complete dissolution of the organic materials. The solids were separated by centrifugation and the solution containing the dissolved organics decanted off. The THF solution was treated with 100 µL of nitrogen-purged BSTFA (N,O-bis(trimethylsilyl)trifluoroacetamide; Fisher Scientific) to form the trimethylsilyl ethers prior to analysis by gas chromatography-mass spectrometry (GC/MS). The reaction mixture was stirred at 80 °C for 60 minutes. The sample was moved to a 1.5 mL GC/MS vial and decane (2µL, internal standard) was added. The mixture was then analyzed by GC/MS. In all GC/MS analyses, an injection volume of 4 µL was used in an Agilent 6890 series GC System equipped with a HP-5MS column and Agilent 5973 Network Mass Selective Detector. The column (30m x 0.25mm ID and film thickness of 0.25 µm, Agilent) was used with helium as the carrier gas at a constant flow rate of 1.0 mL/min. The injector was held at 225 °C. The column temperature program consisted of injection at 50 °C and holding for 10 minutes. The temperature was then ramped to 200 °C at a rate of 10 °C/min and held at 200 °C for 15 minutes for a total run time of 40 minutes. The MS was operated in the electron impact (EI) mode at 70 eV. The scan range was set from 50 to 550 Da. In all cases, the retention time of decane was set to 12.80 min.

Data were acquired and processed with HP-Chemstation software. Quantitative data were obtained by comparing peak heights of the analytes of interest. When analyzing the MS of the sugars and sugar derivatives, it has been shown that the TMS ethers can be unstable under the conditions of ionization and can show many unexpected fragmentations. For example, the $[M]^+$ fragment is almost always missing in these compounds. The expected $[M-CH_3]^+$ (M-15) and $[M-OTMS]^+$ (M-89) are also often not present, but the combined $[M-CH_3-OTMS]^+$ fragment can sometimes be seen (Schummer et al. 2009). Compounds in the shocked samples were identified through comparison of their mass spectra with those of authentic samples. Retention times were used as a secondary identification method in cases where the MS was incomplete due to low abundances of the analyzed compound. The Supplemental Materials contain the complete GC/MS analyses of unshocked and control experiments.

**Results and Discussion**

*Analysis of Unshocked GLA and Montmorillonite Clay*

In order to provide a baseline with which to compare the shocked samples of GLA and GLA/clay mixtures, analyses of the unshocked GLA and unshocked GLA/clay were conducted in parallel with the shocked samples. In the solid phase, GLA exists predominantly in the dimer form. Two dimers have been identified: a five- and a six-membered ring, shown in Figure 1. When the GLA is dissolved, it dissociates to form the monomer with the 5- and 6- membered dimer in equilibrium. The amount of the different species in solution is highly dependent on concentration and temperature and has been investigated elsewhere (Glushonok et al. 2000).

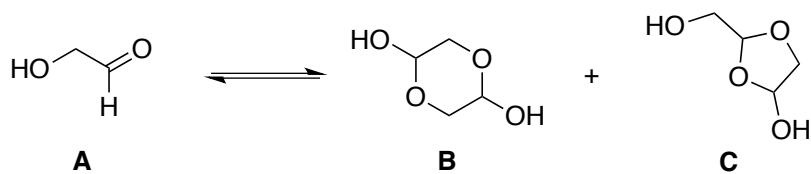

**FIGURE 1.** Structures of GLA monomer (**A**), 6-membered ring dimer (**B**), and 5- membered ring (**C**) dimer present in solution.

When the solid GLA is mixed with montmorillonite clay (unshocked) and held at room temperature or in the freezer at -80 °C, only minor chemical changes can be seen by GC (Figure 2). The predominant component of the sample is the 6-membered ring dimer (B) at 19.26 min. Also present are peaks representing the monomer (A) at 14.68 min and the 5- membered ring dimer (C) at 19.32 min. The five- and six- membered ring dimers were identified by comparison of the MS to published spectra (Novina 1984).

In the unshocked control, four new compounds were seen in the GC. At 15.35 min, a new compounds was identified as glycolic acid through retention time and MS matching. Three new compounds have been identified with retention times longer than 18.5 minutes (inset, Figure 2).

At 18.95 minutes, there is a small peak that has been identified as glycerol through matching of the MS and retention times to an authentic sample. Table 1 shows the amount of glycolic acid and glycerol present in the unshocked control relative to the decane standard. The presence of glycerol was unexpected and it suggests that a series of reactions of GLA are occurring at the clay surface in the solution phase, including dimerization and reduction reactions, but the exact mechanism of the formation of this compound is not known at this time.

**Table 1.** Amount of compound present relative to decane.

|  | **Unshocked GLA/Clay (1:20)** | **3598 4.65 GPa** | **3604 12 GPa** | **3622 25.1 GPa** |
|---|---|---|---|---|
| **Sample Size (mg)** | 27 | 24 | 28 | 25 |
| **Recovery Compared to Unshocked Control** |  | 95% | 96% | 6.5% |
| **Glycerol** | 0.28% | 0.76% | 0.34% | -- |
| **Glycolic acid** | 1.0% | 2.0% | 1.0% | -- |
| **Threose (all forms)** | 0.4% | 3.6% | 1.2% | 1.2% |
| **Erythrose (all forms)** | 0.2% | 0.6% | 0.3% | 0.2% |

Peaks at 20.12 and 20.36 minutes (identified with D and E on Figure 2 inset) have not been identified but the mass spectra of these compounds show fragments at m/z 73, 147 and 204, indicating the presence of hydroxylated compounds. The m/z 73 fragment is from the trimethylsilyl (TMS) ion and is present in all MS of silylated compounds. The m/z 147 and 204 peaks are characteristic of compounds that contain more than one OTMS group (Petersson 1984). The identification of these compounds is currently underway.

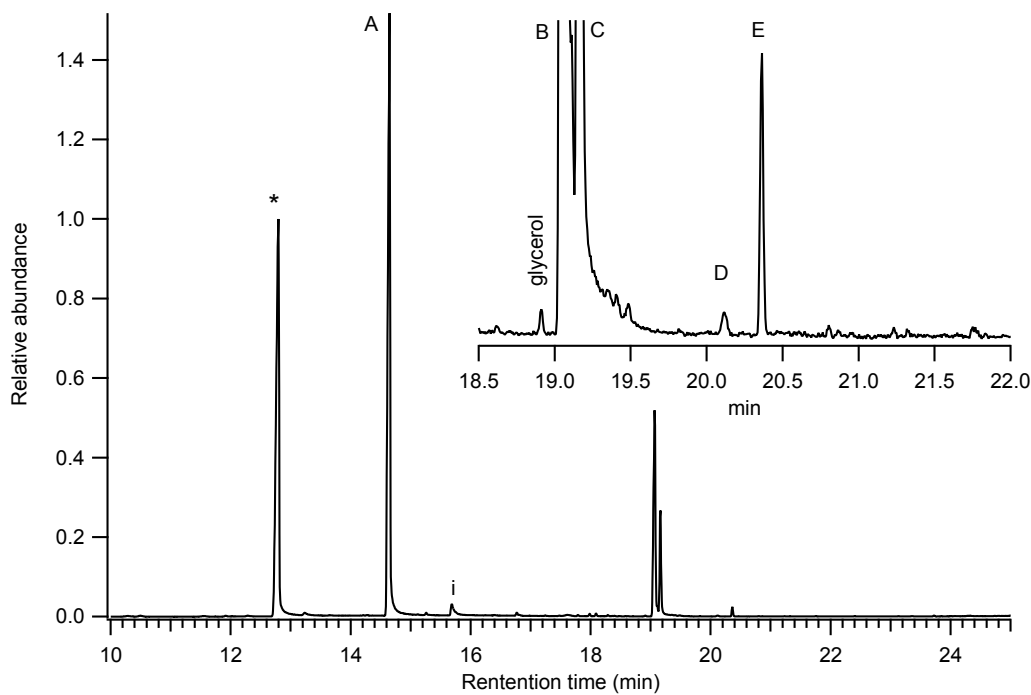

**FIGURE 2.** Gas chromatogram of the GLA mixed with montmorillonite clay. Identity of peaks: (*) decane, (i) impurity from derivatization. Inset shows the expanded region from 18.5 to 22 minutes. See Figure 1 for structures of A, B and C.

*Analysis of Shocked GLA and Montmorillonite Clay*

When the GLA is mixed with montmorillonite clay and subjected to the shock experiments, survival of the sugar was seen at all pressures and additional peaks at longer retention times were observed in the gas chromatograms (Figure 3). Peaks corresponding to the GLA monomer and dimers were also present, with survivability of the GLA above 95% for impact pressures of 12 GPa or less. At the highest pressures, survivability of the GLA dropped to 6.5% relative to the unshocked sample, in line with results of the shock chemistry of PAHs (Mimura and Toyama 2005).

Figure 3 compares the GC results of the shocked GLA/clay with an unshocked GLA/clay as a control (Figure 3A-1). The peak at 20.36 minutes (labeled *i*) can be seen in all samples and results from the solution-phase reaction of the GLA with the surface of the clay. As the GLA and clay are subjected to shocks of increasing pressure, several new peaks can be seen in the GC trace (Figures 3A-2, 3A-3, and 3A-4). Glycerol (18.95 min) has been identified in the 4.65 GPa and 12 GPa shocked samples at amounts greater than what was seen in the control samples, but was not present at all in the highest-pressure experiment of GLA/clay/25.1 GPa. The amount of glycerol present in the shocked samples accounted for less than 1% of the total organics measured by GC (c.f. Table 1).

When the area from 20.5 to 22.0 minutes is expanded (Figure 3B), several new peaks can be seen in the shocked samples. Mass spectra of the peaks showed several common mass fragments including m/z 73, 147, 191 and 218. These fragments are characteristic of the tetroses threose and erythrose (Medeiros and Simoneit 2007). In order to conclusively identify the compounds, samples of authentic D-erythrose and D-threose were derivatized as TMS ethers, and the resulting GC and mass spectra were compared to those of the shocked samples. Because of the difficulty in using MS fragmentation alone to discriminate between threose and erythrose, the retention times of the authentic silylated sugars were used as a secondary identification method (Table S2). Figure 4 shows the GCs of the GLA/clay/4.65 GPa sample (black) with the authentic silylated samples of D-erythrose (red) and D-threose (blue). The two peaks in the GC trace of the pure tetroses are from the anomeric α- and β-furanose forms of the sugars. The third peak in the D-threose sample is a small impurity of the tautomerized ketose form of the sugar.

After establishing the presence of threose, we looked to determine if the stereoisomer erythrose was also present. Several additional peaks can be seen in the authentic erythrose GC trace, but these were determined to be impurities based on the MS fragmentation patterns and the low intensity of GC peaks. The MS of the authentic erythrose shows the expected fragments at m/z 147, 191 and 218, as does the MS of the shocked sample (Figure S10). The retention times of the peak in the shocked sample match those of the authentic sample, but due to the low MS fragment abundances associated with the small GC peaks, the assignment of these new peaks as being erythrose is tentative.

As can be seen in Table 1, the overall ratio of the newly synthesized tetroses relative to decane decreases with increasing shock pressure, from 3.6% at the lowest pressures to 1.2% at both 12 GPa and 25 GPa. When the relative survivability of GLA and the tetroses is compared, the ratio of GLA to the tetroses is quite large (greater than 50:1) at the two lowest pressures. When the pressure of the impact reached 25 GPa, the ratio still favors the GLA, but the ratio drops to 10:1, suggesting that the survivability of the larger sugars is favored over that of the starting material.

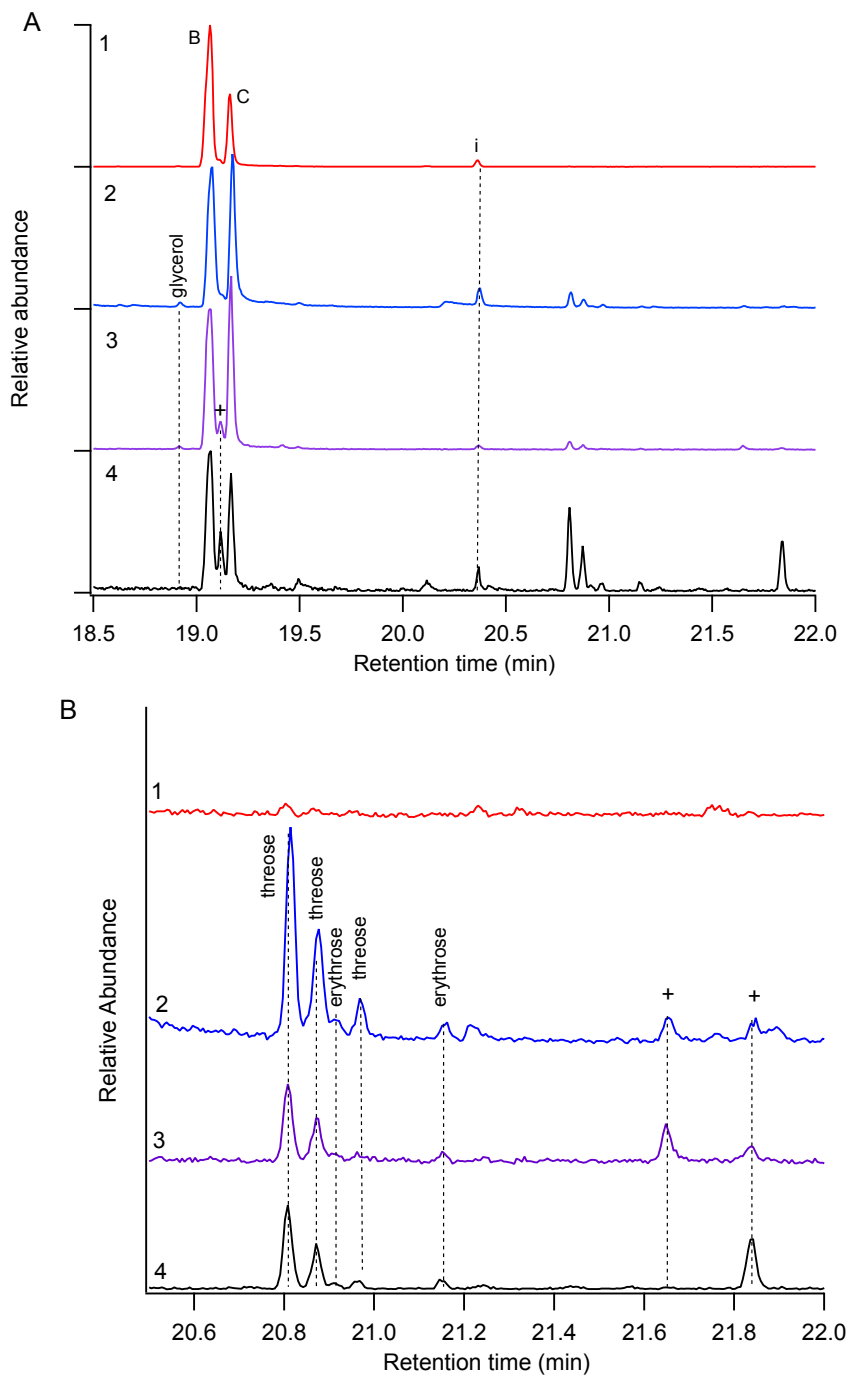

**FIGURE 3.** Gas chromatograms of GLA and montmorillonite clay after impact experiments. 1 is the GLA/clay mixture with no impact. 2, 3, and 4 are samples GLA/clay/4.65 GPa (shot 3598), GLA/clay/12 GPa (shot 3604), and GLA/clay/25.1 GPa (shot 3622), respectively. A) Gas chromatogram of GLA after isolation from the clay. Peaks resulting from the solution phase chemistry of GLA and the clay are labeled with i. Chromatograms are scaled so that the abundance decane is set to 1. B) Expanded region from 20.5 to 22.0 minutes. New, unidentified compounds are labeled with +.

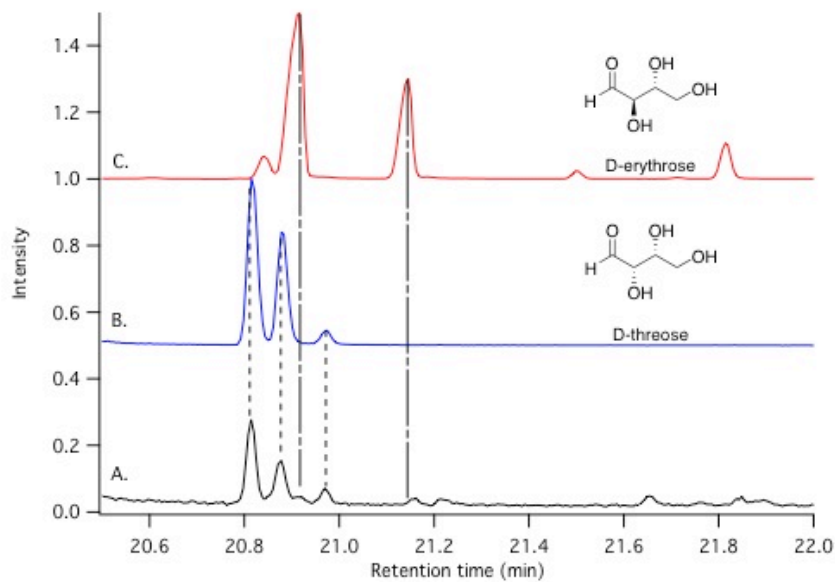

**FIGURE 4**. A. Gas chromatogram of GLA with clay at 4.6 GPa (black). B. The silylated authentic threose with two peaks due to the major α- and β-furanose forms of the sugar, and one at longer retention times to the minor ketose form. C. Silylated authentic erythrose (70% syrup, MP Biomedicals). Peaks at 20.8, 21.5 and 21.8 min are impurities in the sample. The sugars are shown in their extended structures for clarity. See Figures S8-S10 for complete MS analysis of the samples.

The relative ratios of the threose to erythrose being produced in the shocked samples is approximately 9:1 (threose:erythrose) in all experiments. The strong preference for the formation of threose in our experiments was surprising, but a recent review by Cleaves *et al.* (2012) outlines many enantioselective reactions that are catalyzed by minerals and other surfaces. Our results are in contrast to those reported by Lambert *et al.* (2010) where a solution-phase formose reaction of glycolaldehyde favored the formation of erythrose over threose by 3:1. Recent computational studies of the different closed forms of erythrose and threose suggest that the energy differences between the two molecules are minimal (Azofra et al. 2013). The number of energy minima found for D-threose is slightly higher and this result could explain the product distribution in these impact experiments. The observation of threose in these experiments is also intriguing because of the recent postulate that threose nucleic acid (TNA) could have been a chemical progenitor of RNA (Yu et al. 2012).

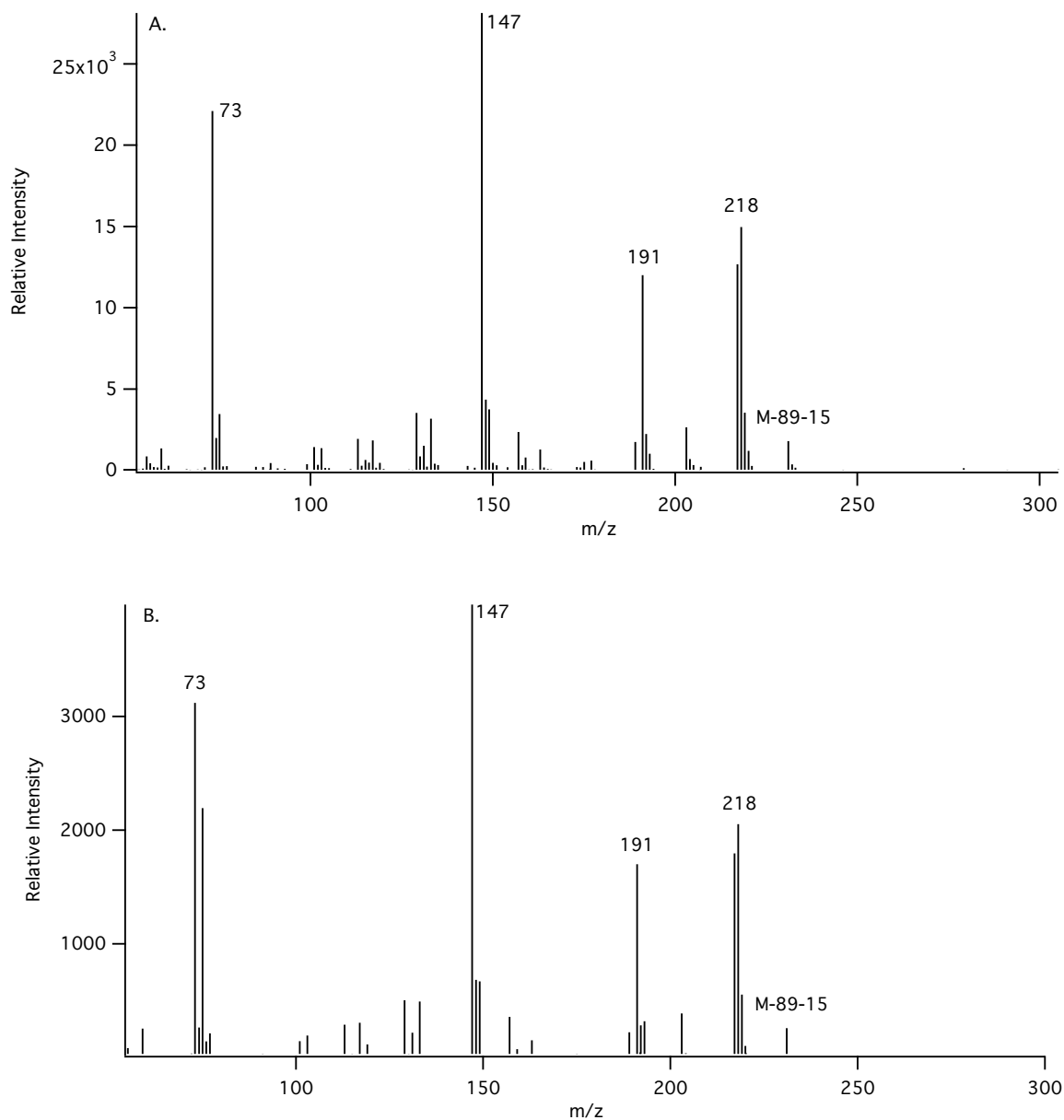

**FIGURE 5**. A. Mass spectrum of the authentic sample of TMS derivatized D-threose (20.79 min). B. Mass spectrum of 20.81 min peak from GLA/clay/4.6 GPa sample (Fig. 4A).

The presence of threose and erythrose in the shocked samples is attributed to the reaction of the GLA with the surface of the montmorillonite clay. Control experiments with neat GLA were performed with comparable shock pressures and in the resulting samples, no evidence of formation of threose or erythrose was seen (Figures S6 and S7). Small peaks from new compounds were seen in the GC but due to the very low abundances of these new peaks, these have not yet been identified. In the neat GLA shock experiments, survivability of the GLA was seen with greater than 90% of the sample recovered at lower pressures (4.6 and 9.4 GPa). At the highest pressure shot (25.6 GPa), the neat GLA did not survive the simulated impact. These results are in agreement with pilot studies conducted at the NASA Ames Vertical Gun (Zellner et al. 2011). The presence of the clay in the shock experiments is necessary for the survival and reaction of the GLA to form larger and more complex sugars.

Additional peaks can be seen in the GLA/clay shocked samples at 21.65 and 21.83 minutes (Figure 3B, indicated with +). Mass spectra of these compounds show fragmentation peaks at m/z 75, 103, 117, 147 and 191, again suggesting that they are TMS derivatives of polyhydroxylated compounds. At the lowest pressure shot, both peaks are present at approximately the same intensity, but at very low abundances compared to those of the identified tetrose peaks. Work is ongoing in our lab to identify these new compounds in the shocked samples.

There were few other peaks seen in the GC at retentions times shorter than 18 minutes that could be attributed to reactions between the GLA and clay during the shock event. Two additional compounds have been identified, but in different samples. In the highest pressure experiment (GLA/clay/25.1 GPa), one peak at 12.51 min was identified as ethylene glycol through matching of the mass spectrum of the shocked sample to an authentic, derivatized sample of ethylene glycol, Figure 6. This new compound results from the reduction of the parent GLA on the clay surface. However, ethylene glycol was not identified in either of the GLA/clay/4.65 GPa or GLA/clay/12 GPa samples, indicating it can only form under conditions of high impact pressures. In both GLA/clay/4.65 GPa and GLA/clay/12 GPa samples, glycolic acid was identified through retention time and MS matching. The concentration relative to decane was low, 2.0% and 1.0% respectively. Other studies have also seen varying product ratios depending on the impact pressures (Blank *et al.* 2001, Mimura and Toyama 2005) and have attributed this to changing reaction mechanisms with the different pressures.

The presence of ethylene glycol in experiments with GLA/clay (Figure 6) can be attributed to the reduction of the parent GLA by the surface of the montmorillonite clay. Although rare, reduced starting materials have been seen before as side products in aldol condensations catalyzed by clays (Azzouz et al. 2003). Additionally, it could be formed through a Cannizaro reaction, but the absence of glycolic acid in the samples containing ethylene glycol suggests that a different mechanism is occurring. The observation of ethylene glycol in these experiments is interesting because of its detection in comet C/1995 O1 (Hale-Bopp; Crovisier et al. 2004) and because of the detection of phyllosilicates in the nucleus of comet 9P/Temple 1 (Lisse et al. 2006). Phyllosilicates are smectite-group minerals, of which montmorillonite is a member. Thus the interaction of GLA and clay on the comet nucleus could be a possible mechanism for the formation of the ethylene glycol. Furthermore, the irradiation of ethylene glycol ices has been shown to readily form GLA (Hudson et al. 2005), which could then be delivered to a young planet via cometary impact.

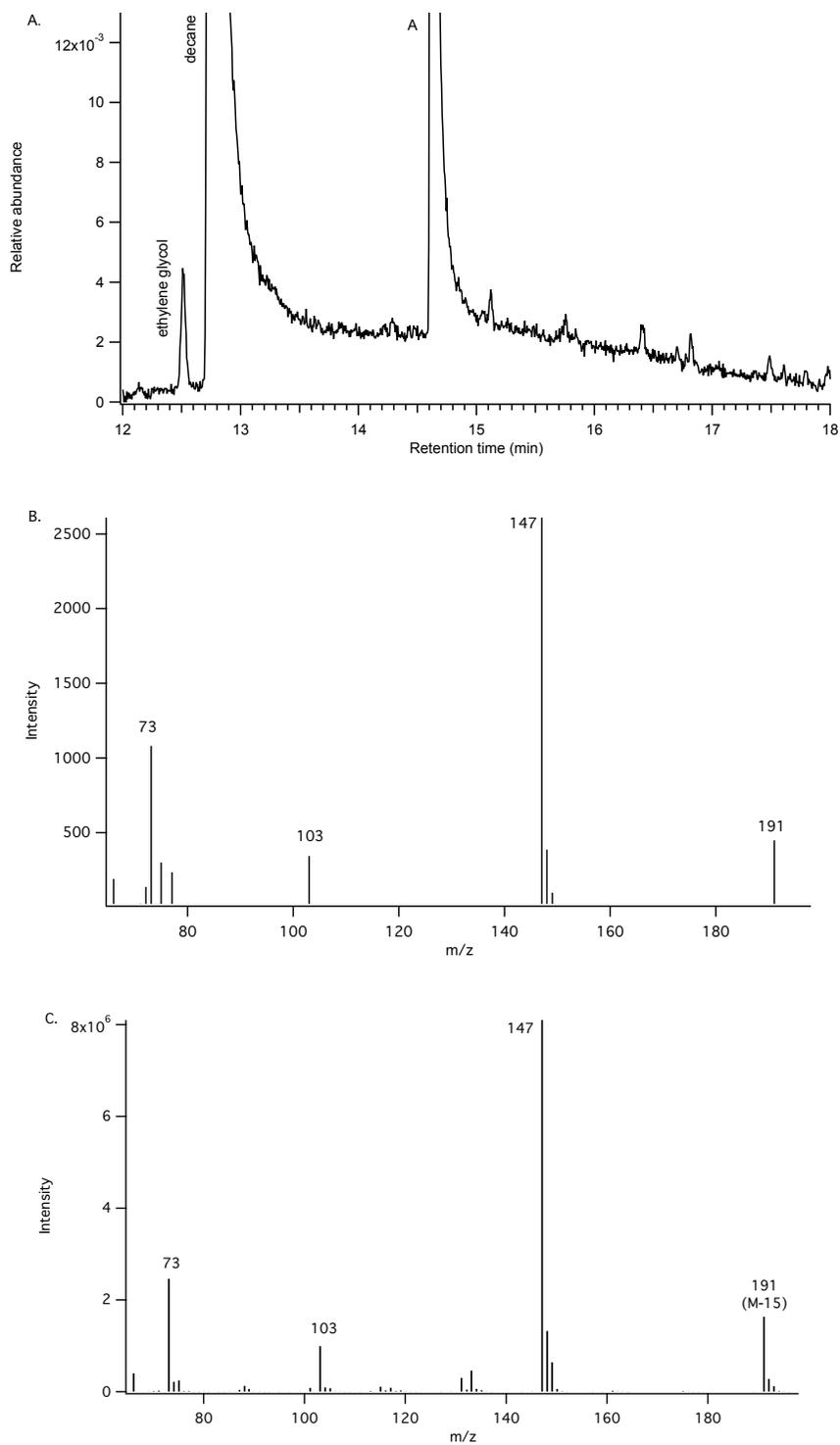

**FIGURE 6.** A.) Gas chromatogram of GLA/clay/25.1 GPa sample, 12-18 minutes. B.) Mass spectrum of 12.51 min peak from shocked sample. C.) Mass spectrum of an authentic sample of the di-TMS derivative of ethylene glycol.

**Conclusions**

The survivability of glycolaldehyde, both neat and mixed with a clay matrix, under conditions that mimic meteoritic impact was assessed. When the impact pressures range from ~5 GPa to >25 GPa, biologically relevant complex sugars are produced. In particular, when mixed with a mineral matrix that was chosen to correspond to volcanic ash that the organic molecule might have encountered upon impact with an early Earth, glycolaldehyde was found to have a high survivability. Several products were conclusively identified, including threose, erythrose, ethylene glycol and glycerol, at varying abundances according to shock pressure. The reactions and changes to molecular structure are most likely due to shock loading and not to prolonged sitting of the target and sample after impact, in agreement with previous investigators (Martins et al. 2013; Burchell et al. 2010; Blank et al. 2001; Peterson et al. 1997). Though other investigators have shown survivability and formation of amino acids, this is the first set of experiments, relevant to comet and/or asteroid impact delivery, to show the survivability and formation of sugars.

In all experiments, there were many new compounds formed in the impact that were not conclusively identified. These new compounds show MS fragmentation patterns consistent with hydroxyl functional groups, but the exact structures are not known at this time. Current work in our lab is focused on the identification of these compounds through comparison with known samples of alcohols and other derivatives, including carboxylic acids and large sugars and sugar alcohols.

Biologically interesting molecules have been observed in space, including in young star systems, and it is likely that these molecules are delivered to Earth (and other planets). While velocity and pressure extrapolations from laboratory experiments to large natural impact events have been described (DeCarli et al. 2002;  Holsapple 1993) and thus allow us to make comparisons between these two extreme conditions, shock pulses and cooling times will be different and may provide substantial limitations for direct comparisons. Nonetheless, results from laboratory experiments can provide a baseline comparison and serve as a way to understand how biologically relevant molecules are both delivered and affected by impacts, especially if those impacts are occurring on young Earth-like planets.

**Acknowledgements**

This project was funded by a grant from the NASA Astrobiology Institute's Directors' Discretionary Fund and, in part, by a grant from the NASA Exobiology and Evolutionary Biology Program (10-EXO10-0109). NEBZ and VPM were additionally supported by grants from the Hewlett Mellon Fund for Faculty Development at Albion College, Albion, MI. ERB, EKE, and CMW were supported by Albion College's Foundation for Undergraduate Research, Scholarship and Creative Activity. We thank Mark Cintala, Frank Cardenas and Roland Montes at the NASA Johnson Space Center's Experimental Impact Facility for their assistance during the impact experiments. We also thank David Carey at Albion College for assistance in the chemistry labs and John Delano at the University of Albany (SUNY) for providing the montmorillonite clay. The authors thank Prakash Joshi, Gavin Reid, John Delano, Bruce Watson, Lisa Lewis and Jim Dye for constructive comments, and VPM thanks Chris Rohlman at Albion College for discussions about biochemistry.  Finally, the authors thank two anonymous reviewers for comments that improved this manuscript.

**References**

Anders E (1989) Prebiotic organic matter from comets and asteroids. *Nature* 342:255-256.

Azofra LM, Alkorta I, Elguero J, Popelier PLA (2013) Conformational study of the open-chain and furanose structure of D-erythrose and D-threose. *Carbohydr Res* 358:96-105.

Azzouz A, Messad D, Nistor D, Catrinescu C, Zvolinschi A, Asafei S (2003) Vapor phase aldol condensation over fully ion-exchanged montmorillonite-rich catalysts. *Appl Catal A* 241:1-13.

Beltrán MT, Codella C, Viti S, Neri R, Cesaroni R (2009) First detection of glycolaldehyde outside the galactic center. *Astrophys J* 690:L93-L96.

Bennett CJ and Kaiser RI (2007) On the formation of glycolaldehyde ($HCOCH_2OH$) and methyl formate ($HCOOCH_3$) in interstellar ice analogs. *Astrophys J* 661:899-909.

Bertrand M, van der Gaast S, Vilas F, Hörz F, Haynes G, Chabin A, Brack A, Westall F (2009) The fate of amino acids during simulated meteoritic impact. *Astrobio* 9:943-951.

Bishop JL, Dobrea EZN, McKeown NK, Parente M, Ehlmann BL, Michalski JR, Millikin RE, et al. (2008) Phyllosilicate diversity and past aqueous activity revealed at Mawrth Vallis, Mars. *Science* 321:830–833.

Blank JG, Miller GH, Ahrens MJ, Winans RE (2001) Experimental shock chemistry of aqueous amino acid solutions and the cometary delivery of prebiotic compounds. *Orig Life Evol Biosph* 31:15-51.

Botta O and Bada JL (2002) Extraterrestrial organic compounds in meteorites. *SGeo* 23:411-467.

Burchell MJ, Parnell J, Bowden SA, Crawford IA (2010) Hypervelocity impact experiments in the laboratory relating to lunar astrobiology. *Earth Moon Planets* 107:55-64.

Burton AS, Glavin DP, Callahan MP, Dworkin JP, Jenniskens P, Shaddad MH (2011) Heterogeneous distributions of amino acids provide evidence of multiple sources within the Almahata Sitta parent body, asteroid 2008 TC3. *MAPS* 46:1703-1712.

Chyba CF, Thomas PJ, Brookshaw L, Sagan C (1990) Cometary delivery of organic molecules to early earth. *Science* 249:366-373.

Cleaves HJ, Scott AH, Hill FC, Leszczynski J, Sahai N, Hazen R (2012) Mineral-organic interfacial processes: potential roles in the origins of life. *Chem Soc Rev* 41:5502-5525.

Cooper GW, Kimmich N, Belisle W, Sarinana J, Brabham K, Garrel L (2001) Carbonaceous meteorites as a source of sugar-related organic compounds for the early Earth. *Nature* 414:879-883.

Crovisier J, Bocklelée-Morvan D, Biver N, Colom P, Despois D, Lis DC (2004) Ethylene glycol in comet C/1995 01 (Hale Bopp). *Astron Astrophys* 418:L35-L38.

DeCarli PS, Bowden E, Jones AP, Price DG (2002) Laboratory impact experiments vs. natural impact events. In *Proceedings of the 2000 Vienna Conference on Catastrophic Events and Mass Extinctions: Impacts and Beyond*, eds. C. Koeberl and K. MacLeod. *GSA Special Paper* 356:595-605.

Delano JW, Tailby ND, Aldersley MF, Watson EB, Joshi PC, and Ferris JP (2010) Could montmorillonites have played a role in the formation of prebiotic molecules on the early Earthy? [abstract 2525]. In 41[st] *Lunar and Planetary Science Conference Abstracts*, Lunar and Planetary Institute, Houston.

Elsila JE, Glavin DP, Dworkin JP (2009) Cometary glycine detected in samples returned by Stardust. *MAPS* 44:1323-1330.

Farley KA, Malespin C, Mahaffy P, Grotzinger JP, Vanconcelos PM, et al. (2014) In Situ Radiometric and Exposure Age Dating of the Martian Surface. *Science* 343:1247166-1-5.

Ferris JP (1998) Catalyzed RNA synthesis for the RNA world. In Molecular Origins of Life, p. 255-268. Cambridge Univ. Press.

Ferris JP (2005) Mineral catalysis and prebiotic synthesis: Montmorillonite-catalyzed formation of RNA. *Elements* 1:145-149.

Ferris, JP and Ertem G (1993) Montmorillonite catalysis of RNA oligomer formation in aqueous solution. A model for the prebiotic formation of RNA. *J Am Chem Soc* 115:12270-12275

Ferris JP and Hagan WJ (1986) The adsorption and reactions of nucleotides on montmorillonite clays. *Orig Life Evol Biosph* 17:69-84.

Flynn GJ and McKay DA (1990) An assessment of the meteoritic contribution to the martian soil. *J Geophys Res* 95 (B9):14497.

Gibbons RV (1974) *Experimental Effects of High Shock Pressure on Materials of Geological and Geophysical Interest*. Ph.D. Thesis, California Institute of Technology (Pasadena, CA), pp 226.

Gibbons RV and Ahrens TJ (1971) Shock metamorphism of silicate glasses. *J Geophys Res* 76:5489-5498.

Glavin DP, Elsila JE, Burton AS, Callahan MP, Dworkin JP, Hilts RW, Herd CDK (2012) Unusual nonterrestrial L-proteinogenic amino acid excesses in the Tagish Lake meteorite. *MAPS* 47:1347-1364.

Glushonok GK, Glushonok TG, Shodryo OI (2000) Kinetics of equilibrium attainment between molecular glycolaldehyde structures in an aqueous solution. *Kinet Catal* 41:620-624.

Gross JH (2011) in *Mass Spectrometry, 2$^{nd}$ edition*, Chapter 2, Springer.

Halfen DT, Apponi AJ, Woolf N, Polt R, Ziurys LM (2006) A systematic study of glycolaldehyde in Sagittarius B2(N) at 2 and 3 mm: Criteria for detecting large interstellar molecules. *Astron J* 639:237-245.

Harman CE, Kasting JF, Wolf ET (2013) Atmospheric production of glycolaldehyde under hazy prebiotic conditions. *Orig Life Evol Biosph* 43:77-98.

Hollis JM, Lovas FJ, and Jewell PR (2000) Interstellar glycolaldehyde: The first sugar. *Astrophys J* 540:L107-L110.

Hollis JM, Vogel SN, Snyder LE, Jewell PR, and Lovas FJ (2001) The spatial scale of glycolaldehyde in the galactic center. *Astrophys J* 554:L81-L85.

Hollis JM, Jewell PR, Lovas FJ, Remijan A (2004) Green bank telescope observations of interstellar glycolaldehyde: Low-temperature sugar. *Astrophys J* 613:L45-L48.

Hörz F (1970) A small ballistic range for impact metamorphism studies. NASA TN D-5787.

Holsapple KA (1993) The scaling of impact processes in planetary sciences. *Annu Rev Earth Planet Sci* 21:333-373.

Hudson RL, Moore MH, Cook AM (2005) IR characterization and radiation chemistry of glycolaldehyde and ethylene glycol ices. *Adv Space Res* 36:184-189.

Jalbout A F, Abrell L, Adamowicz L, Polt R, Apponi AJ, Ziurys LM (2007) Sugar synthesis from a gas-phase formose reaction. *Astrobiology* 7:433-442.

Jørgensen JK, Favre C, Bisschop SE, Bourke TL, van Dishoeck EF, Schmalzl M (2012) Detection of the Simplest Sugar, Glycolaldehyde, in a Solar-type Protostar with ALMA. *Astrophys J* 757:L4-L16.

Joshi PC, Aldersley MF, Delano JW, Ferris JP (2009) Mechanism of Montmorillonite Catalysis in the Formation of RNA Oligomers. *J Am Chem Soc* 131:13369-13374.

Keller LP and Zolensky ME (1991) Clay minerals in primate meteorites and interplanetary dust II. Smectites and Micas. Program and Abstracts for Clay Minerals Society, 28th Annual Meeting. Held October 5-10, 1991, in Houston, TX. Hosted with National Aeronautics and Space

McCaffrey/Zellner 21

Administration Lyndon B. Johnson Space Center, and the Lunar and Planetary Institute. LPI Contribution 773, published by the Lunar and Planetary Institute, 3303 Nasa Road 1, Houston, TX 77058, 1991, p.87.

Kieffer, SW (1971) Shock metamorphism of the Coconino sandstone at Meteor Crater, Arizona. *J Geophys Res* 76:5449-5473.

Kim H, Ricardo A, Illangkoon HI, Kim MJ, Carrigan MA, Frye F, Benner SA (2011) Synthesis of carbohydrates in mineral-guided prebiotic cycles. *J Am Chem Soc* 133:9457-9468.

Lambert JB, Gurusamy-Thangavelu SA, Ma K (2010) The silicate-mediated formose reaction: bottom-up synthesis of sugar silicates. *Science* 327:984-986.

Lisse CM, VanCleve J, Adams AC, A'Hearn MF, Fernández YR, *et al.* (2006) Spitzer spectral observations of the Deep Impact ejecta. *Science* 313:635-640.

Love SG and Brownlee D E (1993) A direct measurement of the terrestrial mass accretion rate of cosmic dust. *Science* 262 (5133): 550–553.

MacKinnon IDR and Kaser SA (1988) The clay-size fraction of CI chondrites Alais and Orgueil: An AEM study, *XIX Lun Plan Sci Conf* 19:709.

Martins Z, Price MC, Golman N, Sephton MA, Burchell MJ (2013) Shock synthesis of amino acids from impacting cometary and icy planet surface analogues. *Nature GeoSci* Advanced on-line publication. doi:10.1038/NGEO1930.

Marty B and Yokochi R (2006) Water in the Early Earth. *Reviews in Mineralogy & Geochemistry* 62:421-450.

Medeiros PM and Simoneit BRT (2007) Analysis of sugars in environmental samples by gas chromatography–mass spectrometry. *J Chrom A* 1141:271-278.

Mimura K and Toyama S (2005) Behavior of polycyclic aromatic hydrocarbons at impact shock: Its implication for survival of organic materials delivered to the early Earth. *Geochim Cosmochim Acta* 69:201-209.

Novina R (1984) Determination of the structure of glycolaldehyde as their trimethylsilyl derivatives by gas chromatograph – mass spectrometry. *Chromatographia* 18:96-98.

Owen TC (1998) The origin of the atmosphere. In: Brack A (ed) The molecular origins of life: assembling pieces of the puzzle. Cambridge Press, Cambridge.

Papke K (1969) Montmorillonite Deposits in Nevada, *Clays and Clay Minerals*, 17:211-222.

Pasek M and Lauretta D (2008) Extraterrestrial flux of potentially prebiotic C, N, and P to the Early Earth. *Orig Life Evol Biosph* 38:5-21.

Peterson E, Hörz F, Chang S (1997) Modification of amino acids at shock pressures of 3.5 to 32 GPa. *Geochim Cosmochim Acta* 61:3937-3950.

Petersson G (1984) Mass spectrometry of alditols as trimethylsilyl derivatives. *Tetrahedron* 25:4437-4443.

Pizzarello S, Krishnamurthy RV, Epstein S, Cronin JR (1991) Isotopic analyses of amino acids from the Murchison meteorite. *Geochim Cosmochim Acta* 55:905-910.

Rietmeijer FJM and MacKinnon IDR (1985) Layer silicates in a chondritic porous interplanetary dust particle. *JGR Solid Earth* 90:149-155.

Schummer C, Delhomme O, Appenzeller BMR, Wennig R, Millet M (2009) Comparison of MTBSTFA and BSTFA in derivatization reactions of polar compounds prior to GC/MS analysis, *Talanta* 77:1473-1482.

Shirley JH, Kamp LW, and Dalton JB (2013) Phyllosilicates and Cometary Impacts on the Surface of Europa. AGU Fall Meeting #P54A-07.

Sorrell WH (2001) Origin of amino acids and organic sugars in interstellar clouds. *Astrophys J* 555:L129-L132.

Stephenson JD, Hallis LJ, Nagashima K, and Freeland SJ (2013) Boron Enrichment in Martian Clay. *Plos 1.* doi: 10.1371/journal.pone.0064624.

Stöffler D (1972) Deformation and transformation of rock-forming minerals by natural and experimental shock processes. I. Behavior of minerals under shock compression. *Fortschr Miner* 49:50-113.

Sun VZ and Milliken RE (2014) The Geology and Mineralogy of Ritchey Crater, Mars: Evidence for Post-Noachian Clay Formation. *JGR Planets.* doi: 10.1002/2013JE004602.

Woods PM, Kelly G, Viti S, Slater B, Brown WA, Puletti F, Burke DJ, Raza Z (2012) On the formation of glycolaldehyde in dense molecular cores. *Astrophys J* 750:19-26.

Wozniakiewicz PJ, Ishii HA, Kearsley AT, Burchell MJ, Bradley JP, Teslich N. and Cole MJ (2010) Survivability of cometary phyllosilicates in Stardust collections and implications for the nature of comets. 41[st] Lun Plan Sci Conf, 2357.pdf.

Wray JJ, Ehlmann BL, Squires S W, Mustard JF, Kirk RL (2008) Compositional stratigraphy of clay-bearing layered deposits at Mawrth Vallis, Mars. *Geophys Res Lett* 35:L12202.

Yu H, Zhang S, Chaput JC (2012) Darwinian evolution of an alternative genetic system provides support for TNA as an RNA progenitor. *Nature Chem* 4:183–187.

Zahnle K and Sleep NH (2006) Impacts and the early evolution of life. In: Thomas PJ, Chyba CF, Hicks RD, McKay CP (eds) Comets and the origin and evolution of life. Springer-Verlag, Berlin.

Zellner NEB, McCaffrey VP, Bennett E, Waun C (2012) Assessing the survival of glycolaldehyde after high velocity impacts: Initial experiments and results. In *Proceedings of the 11th Australian Space Science Conference*, eds. Wayne Short and Iver Cairns, Canberra. 26 – 29 September. Published by the National Space Society of Australia Ltd. ISBN 13: 978-0-9775740-5-6.

Zolensky M and Keller LP (1991) Clay minerals in primate meteorites and interplanetary dust I. Program and Abstracts for Clay Minerals Society, 28th Annual Meeting. Held October 5-10, 1991, in Houston, TX. Hosted with National Aeronautics and Space Administration Lyndon B. Johnson Space Center, and the Lunar and Planetary Institute. LPI Contribution 773, published by the Lunar and Planetary Institute, 3303 Nasa Road 1, Houston, TX 77058, 1991, p.184.